\documentclass[aps,10pt,pra,twocolumn,floatfix,floats,showpacs,superscriptaddress,raggedbottom]{revtex4-1}
\usepackage{graphicx,latexsym}
\usepackage{dcolumn}
\usepackage{amsmath}
\usepackage{amssymb,bm}
\usepackage{braket}
\usepackage{natbib}

\newcommand{\angstrom}{\text{\normalfont\AA}}

\usepackage[hidelinks=true,breaklinks=true]{hyperref}
\hypersetup{
  colorlinks   = true, 
  urlcolor     = blue, 
  linkcolor    = blue, 
  citecolor    = red 
}

\usepackage[normalem]{ulem}

\def\sec#1{Sec.\ \ref{#1}}

\def\fig#1{Fig.\ \ref{#1}}

\begin{document}

\title{Silicon on a graphene nanosheet with triangle- and dot-shape: Electronic structure, 
        specific heat, and thermal conductivity from first-principle calculations}

\author{Hunar Omar Rashid}        
\affiliation{Division of Computational Nanoscience, Physics Department, College of Science, 
             University of Sulaimani, Kurdistan Region, Iraq}       
        
\author{Nzar Rauf Abdullah}
\email{nzar.r.abdullah@gmail.com}
\affiliation{Division of Computational Nanoscience, Physics Department, College of Science, 
             University of Sulaimani, Kurdistan Region, Iraq}
\affiliation{Komar Research Center, Komar University of Science and Technology, Sulaimani, Iraq}
 \affiliation{Science Institute, University of Iceland,
        Dunhaga 3, IS-107 Reykjavik, Iceland}

\author{Vidar Gudmundsson}   
\email{vidar@hi.is}
 \affiliation{Science Institute, University of Iceland,
        Dunhaga 3, IS-107 Reykjavik, Iceland}
%

\begin{abstract}
The electronic structure, specific heat, and thermal conductivity of silicon embedded in a monolayer 
graphene nanosheet are studied using Density Functional Theory. Two different shapes of the 
substitutional Si doping in the graphene are studied, a triangular and a dot shape.
The silicon doping of a graphene nanosheet, with the silicon atoms arranged in a triangular configuration 
in ortho- and para-positions, opens up a band gap transforming the sheet to
a semiconducting material.
The opening of the band gap is caused by the presence of the repulsion force between the silicon 
and carbon atoms decreasing the density of states around the Fermi energy. 
Consequently, the specific heat and the thermal conductivity of the system are suppressed.
For graphene nanosheet doped with a dot-like configuration of silicon atoms, at the ortho-, meta-, and para-positions, the valence band crosses the Fermi level. 
This doping configuration increases the density of state at the Fermi level, but
mobile charge are delocalized and diminished around the silicon atoms.
As a result, the specific heat and the thermal conductivity are enhanced.
Silicon substitutionally doped graphene nanosheets may be beneficial for photovoltaics and can further 
improve solar cell devices by controlling the geometrical configuration of the underlying atomic systems.
\end{abstract}



\maketitle

%
%

\section{Introduction}\label{Sec:Introduction}

A graphene nanosheet is a 2D material with remarkable qualities in terms of mechanical \cite{Lee385}, electrical \cite{Zhang2005}, 
chemical \cite{doi:10.1063/1.3676276}, optical \cite{doi:10.1021/nl102824h}, and thermal \cite{Shahil2012} properties.
The investigation of a single layer graphene and its characteristics \cite{Novoselov666} has paved the way 
to generate enormous interest and intense activity in graphene research \cite{Geim2007,doi:10.1021/cr900070d}.
The graphene material has been used as the basic building block for graphitic materials with 
different directions. It may be wrapped up into zero-dimensional fullerene \cite{Kroto1985, doi:10.1002/anie.199201113} leading to 
the improvement of the band gap \cite{Jalali-Asadabadi2016}, rolled into one-dimensional nanotubes \cite{Al-Saleh2015}, 
or stacked into three-dimensional graphite \cite{Feng2014}. 
According to the atomic arrangement, the graphene structures can be classified into two 
categories which are ``zigzag'' and ``armchair''. These two types of graphene have different 
electronic characteristics, especially the creation of non-bonding edge states localized in the zigzag-shaped edges
and electron wave interference in the armchair-shaped ones play important roles in the functionality of graphene \cite{Enoki2012}.
The two types, ``zigzag'' and ``armchair'', can be constructed in the form of graphene nanoribbons which are  
promising structures in electron transport \cite{PhysRevLett.104.056801,Cresti2008,Felix2018}.

The physical properties of graphene can be controlled by doping, which is the process of adding impurities to intrinsic graphene. 
For instance, silicon, Si, doped graphene has emerged as new 2D materials called siligraphenes,
demonstrating attractive optical properties and extreme thermal stability \cite{Li2014,Dong2016}. 
The band gap of siligraphenes depends on the ratio of the Si doping, that is determined by the  
relationship between the reactants and products in a chemical reaction producing the graphene \cite{doi:10.1021/nl403010s,Pumera2017}. One can expect to use g-SiC$_2$ for solar cell materials due to the opening band gap, $1.09$~eV \cite{doi:10.1021/nl403010s}. Increasing the ratio of Si doping to construct
g-SiC$_3$ and g-SiC$_5$, the characteristics of a topological insulator appear in g-SiC$_3$ 
\cite{PhysRevB.89.195427} and g-SiC$_5$ emerges as a semi-metal with excellent gas sensing 
properties \cite{Dong2017}. Furthermore, in g-SiC$_7$ the band gap is increased to $1.13$ eV actively encouraging
photovoltaics devices within the visible light range \cite{Dong2016}.

Recently, there has been an increased and strong motivation to explore thermal characteristics 
of graphene and related composite materials from the technological point of view. 
Electrical and thermal measurements of siligraphenes have shown that g-SiC can be seen as 
fascinating material with interesting properties \cite{Houmad2019}.
It has been demonstrated that the thermal properties of g-SiC$_3$ are better compared to
g-SiC$_7$, and the thermal conductivity of g-SiC$_7$ is exponentially enhanced with
temperature but for g-SiC$_3$ it is parabolically changed \cite{Houmad2018}.
Exploration has shown that the effective thermal conductivity in an optimized mixture of
graphene and multilayer graphene can be enhanced \cite{Shahil2012}.
Furthermore, both graphene and graphite at room temperature can be utilized to increase the efficiency 
of solar cell devices due to a high-recorded thermal property dominated by the acoustic phonons \cite{Balandin2011}.

Motivated by the aforementioned studies, we model a graphene and siligraphene nanosheets. The 
electronic and the thermal characteristics are studied using Density Functional Theory (DFT).
We model two different shapes of substitution Si doping in the graphene: triangle and ``dot'' shapes.
In the triangle shape the Si-atoms are substitutionally embedded in the ortho- and para-positions of 
the honeycomb structure of graphene. This opens up a band gap leading to
the suppression of thermal conductivity. In the ``dot'' shape, the Si-atoms are substitutionally 
doped in the ortho-, para-, and meta-positions of 
graphene. In the ``dot'' structure, the enhancement of thermal properties of the system is observed.

In \sec{Sec:Model} the structure of graphene nanosheet is briefly overviewed. In \sec{Sec:Results} the main achieved results are analyzed. In \sec{Sec:Conclusion} the conclusion of the results is presented.

\section{Model}\label{Sec:Model}

\begin{figure}[htb]
\centering
\includegraphics[width=0.15\textwidth, angle =270 ]{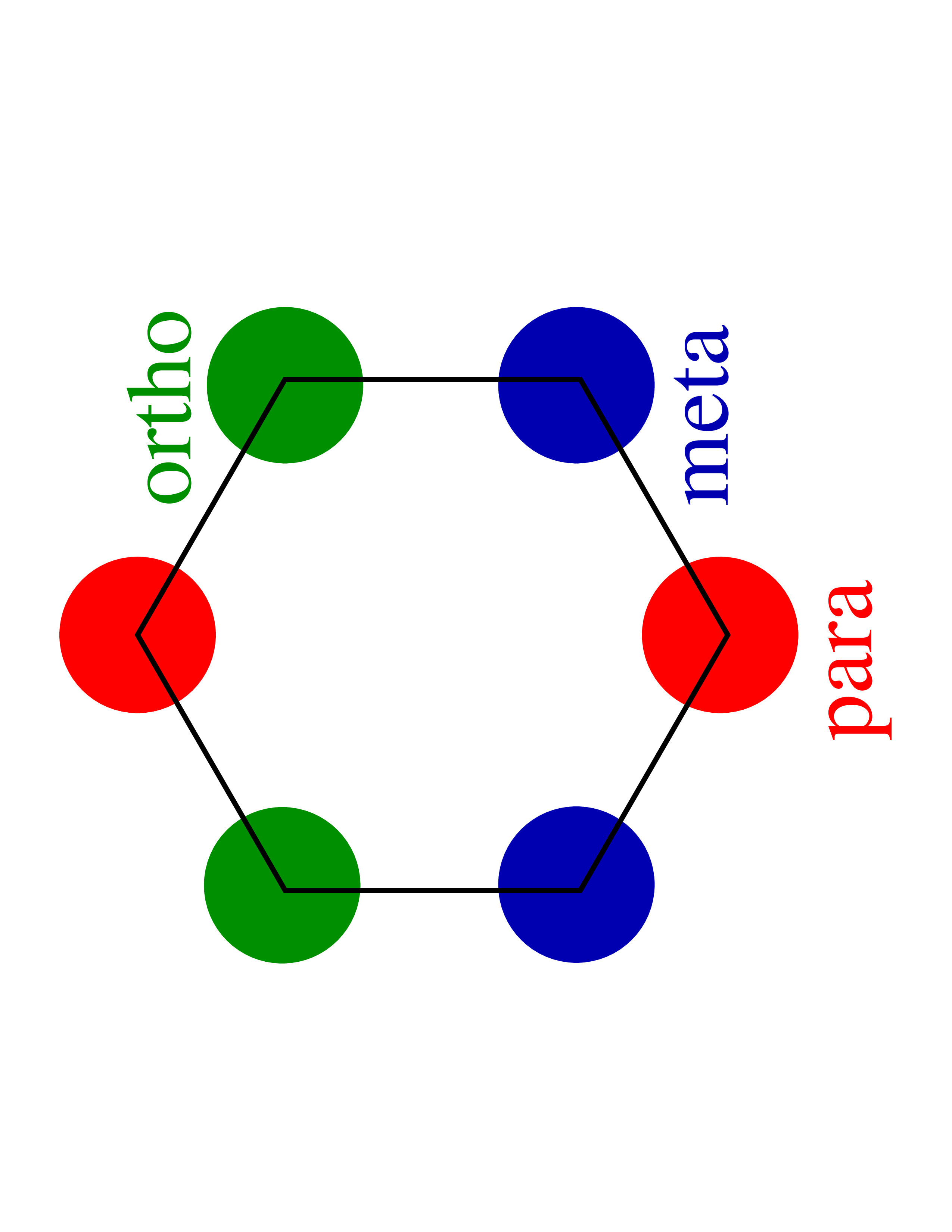}
 \caption{Schematic diagram indicating the ortho (green), the meta (blue), and the para (red) positions 
 of doping in the honeycomb structure.}
\label{fig01}
\end{figure}

We model a monolayer graphene nanosheet consisting of a $3\times3$ 
supercell with a  diamond shape that is comprised from $32$ carbon atoms. 
We consider the vacuum space in $z$-axis to be $9.74$ ${\angstrom}$. 
The convergence of the SCF calculation is set to $10^{-3}$ eV, and the geometry of the system is fully relaxed with a Gamma-centered $8\times8\times1$ k-mesh 
for both pure and doped graphene nanosheets until the calculated force is smaller than $0.008$ eV/${\angstrom}$. 
In addition to the pristine graphene, we consider two geometrical shapes of Si-atoms in doped graphene, the  
triangle and the ``dot'' shape. The triangle Si doped graphene is formed
if two Si-atoms are put at the ortho-positions (green) and one Si-atom is placed in a para-position (red) 
as is shown in \fig{fig01}. The ``dot'' Si doped graphene can be built by adding two Si-atoms at the ortho-positions, 
two Si-atoms at the meta-positions (green) and two Si-atoms at the para-positions \cite{Rani2014}.

The electronic structure is calculated via the plane-wave projector-augmented wave
method implemented in the Quantum Espresso (QE) package \cite{Giannozzi_2009}. In the QE package, 
the approach is based on an iterative solution of the Kohn-Sham equation of the DFT theory 
\cite{PhysRev.140.A1133}. 
In the DFT approach, the generalized gradient approximation (GGA) method, and the exchange-correlation 
functions are realized in the non-relativistic Perdew-Burke Emzerhof pseudo-potential (PBE) \cite{Petersen2000}.
In addition, The plane-wave basis is arranged to a kinetic energy cut-off equal to $490$ eV 
\cite{PhysRevLett.77.3865}. The DFT scheme can thus be used to investigate the band structure, the density 
of state (DOS), and the charge density distribution \cite{Vidar:ANDP201500298} of the system.

The thermal properties of the system 
are studied using the Boltzmann theory implemented in the BoltzTraP package \cite{Madsen2006},
where the specific heat, $c$, of the system can be calculated via 
\begin{equation}
 c(T;\mu) = \int n(\varepsilon) (\varepsilon - \mu) \Big[   \frac{\partial f_{\mu}(T;\varepsilon)}{\partial T}\Big] \, d\varepsilon
\end{equation}
and the electronic thermal conductivity, $\kappa^0$, is determined by
\begin{equation}
 \kappa_{i,j}^0(T;\mu) = \frac{1}{e^2 T \Omega} \int \sigma_{i,j}(\varepsilon) (\varepsilon - \mu)^2 \Big[   -\frac{\partial f_{\mu}(T;\varepsilon)}{\partial \varepsilon }\Big] \, d\varepsilon
\end{equation}
where $\sigma_{i,j} (\varepsilon)$ indicates the conductivity tensors, and 
$\Omega$ is the number of $K$ point which are sampled in Brillouin zone  \cite{Madsen2006}.

\section{Results}\label{Sec:Results}

\begin{figure}[htb]
\centering
    \includegraphics[width=0.23\textwidth]{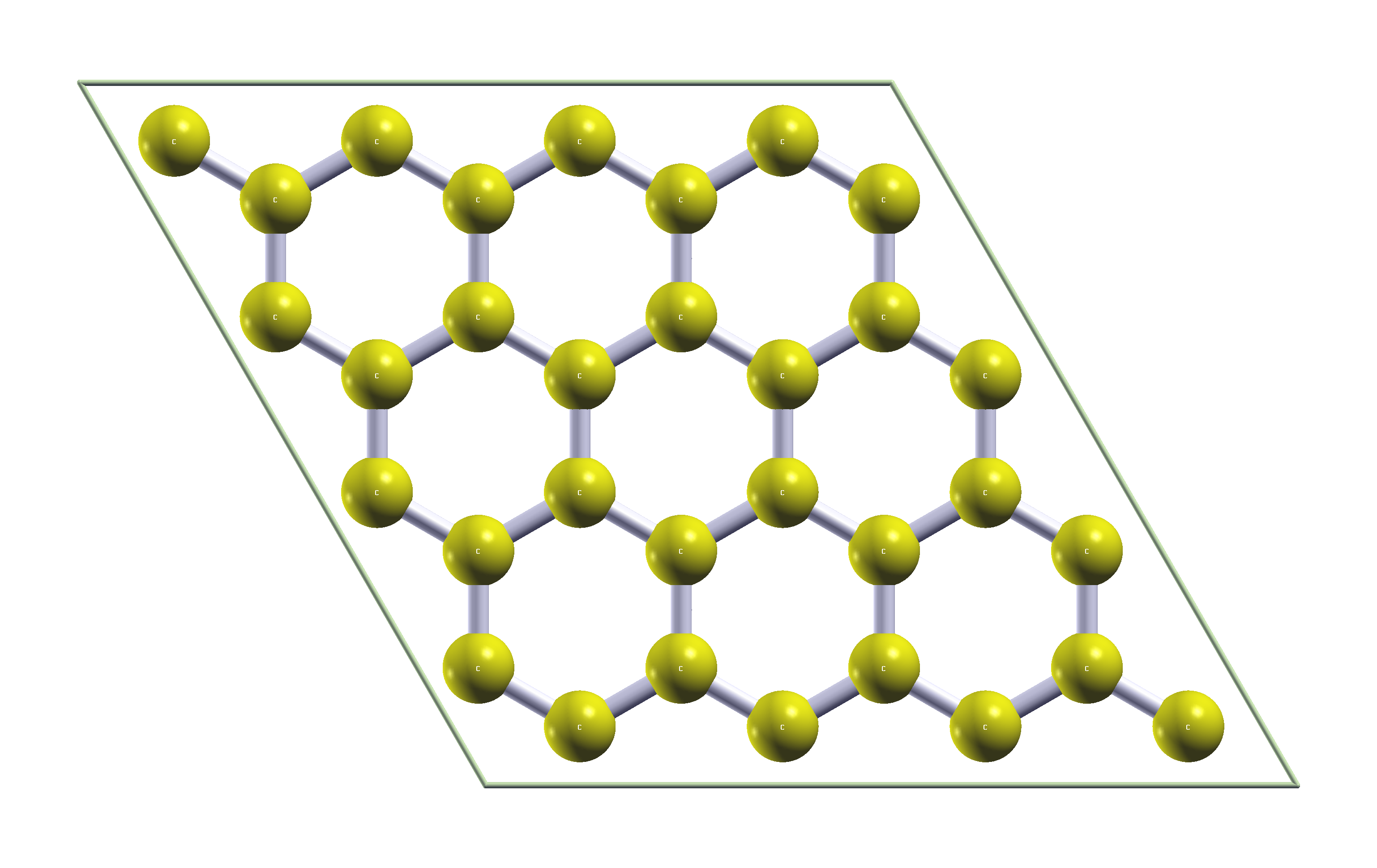}
    \includegraphics[width=0.23\textwidth]{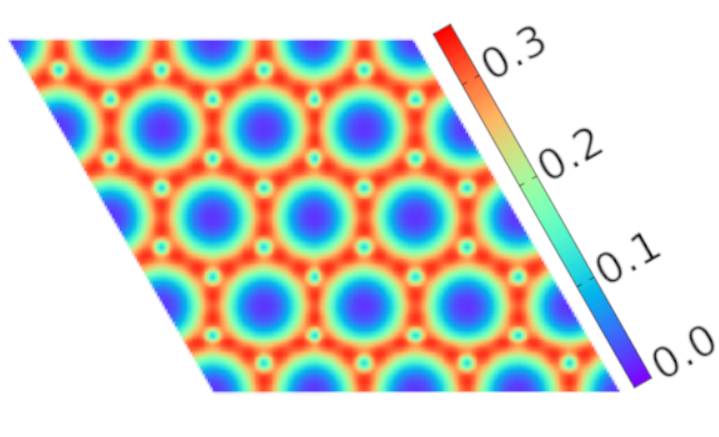}
 \caption{The pure graphene nanosheet with $3\times3$ supercell (left panel) and the charge density distribution (right panel).
          The carbon atoms, C, are golden colored.
          The bond length C-C is $1.42$ ${\angstrom}$, and the lattice constant is $a = 2.46$ ${\angstrom}$.
          Xcrysden is used to produce the pure graphene supercell (left panel) \cite{KOKALJ2003155}.}
\label{fig02}
\end{figure}

In this section, we present the main results obtained from the calculations.
We start with the pristine graphene nanosheet without an Si-dopant. 
Initially, we let the system fully relax. After the structural relaxation, the bond length of the C-C atoms 
is found to be $1.42$ ${\angstrom}$, and the lattice constant becomes $a = 2.46$ ${\angstrom}$, 
these values are in good agreement with the literature \cite{Avouris2010}. 
Figure \ref{fig02} displays the pristine graphene nanosheet (left panel) and 
its charge density distribution (right panel).  
It seems that the honeycomb structure with the $3\times3$ supercell is clearly observed in 
the charge density distribution without any defect or deformation in the crystal structure 
indicating a pure graphene nanosheet.

\begin{figure}[htb]
\centering
\includegraphics[height=0.3\textwidth]{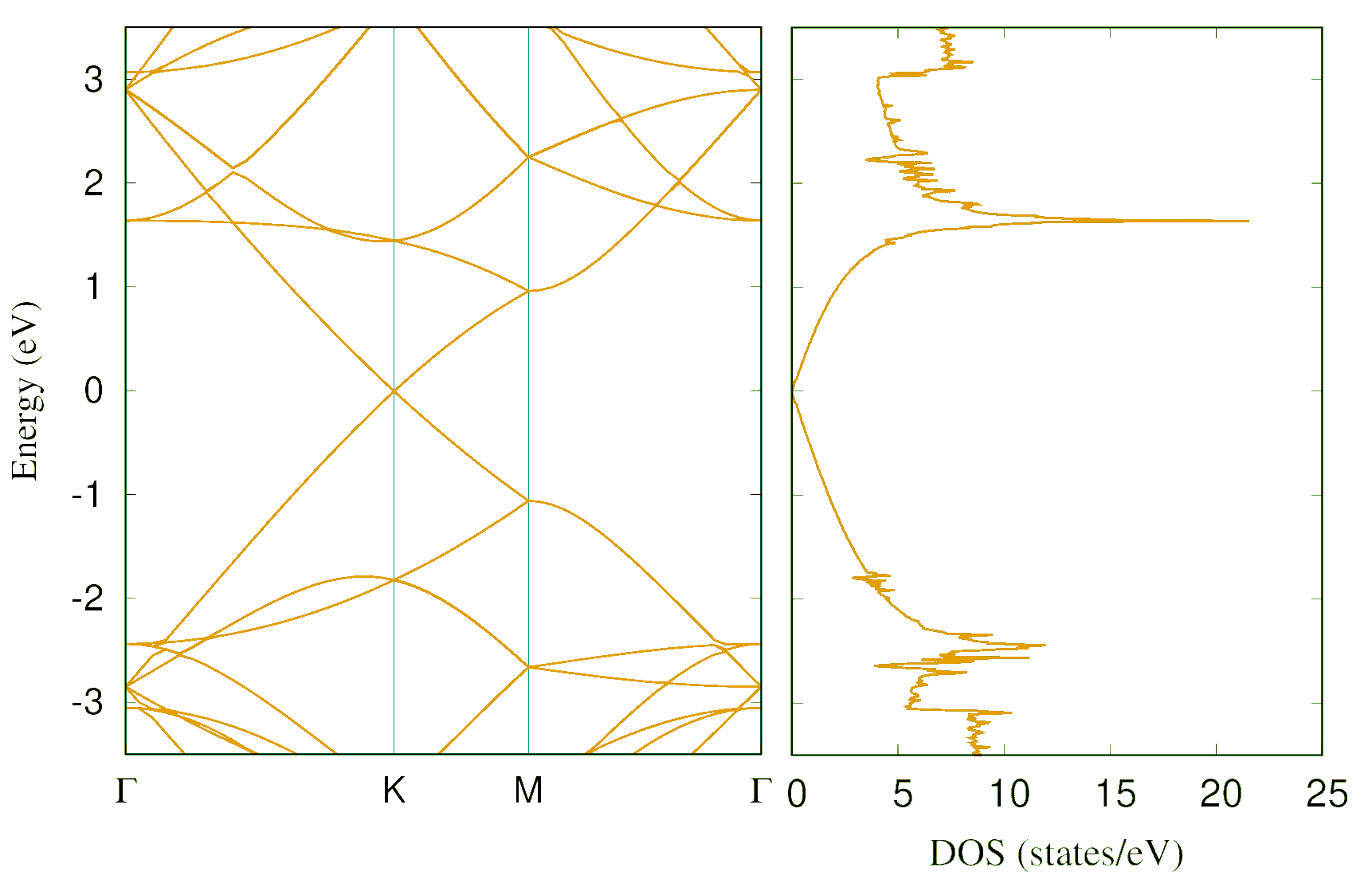}
 \caption{Band structure (left panel) and the DOS (right panel) of the pristine monolayer graphene nanosheet.
        The horizontal dashed black line represents the Fermi level, $E_{F}$. The bond length C-C is 
        $1.42$ ${\angstrom}$, and the lattice constant is $a = 2.46$ ${\angstrom}$.}
\label{fig03}
\end{figure}

The electronic structure of the pristine graphene system is presented in \fig{fig03}, 
with the band structure (left panel) and the density of state (right panel). 
The dashed black line in the energy axis indicates the Fermi level, $E_{F}$.
As expected, there is no band gap between the valence and the conduction bands at the $K$ point. 
It turns out that the DOS is zero at the point where the band gap is zero (see \fig{fig03} (right panel)).
The band structure and the DOS of pristine graphene for different numbers of supercells 
have been investigated by many research groups, where the zero band gap and the DOS 
have been predicted for a monolayer \cite{PhysRevB.77.045407}, a bilayer \cite{Zollner_2018}, and a trilayer  
\cite{Li2014,doi:10.1063/1.4902846}.
We note that in our calculations the spin-orbit interaction (SOI) is neglected. 
In the presence of the SOI, we find a tiny band gap at the $K$ point with a magnitude $0.98 \, \mu{\rm eV}$ which 
is in good agreement with an estimate obtained from a tight-binding model \cite{PhysRevB.75.041401}. 
The gap can be referred to the interactions of the $\pi$ orbital bonds.
A bit larger band gap is seen when higher orbits of the carbon atoms are included in the calculations 
\cite{PhysRevB.80.235431}.

\begin{figure}[htb]
\centering
    \includegraphics[width=0.23\textwidth]{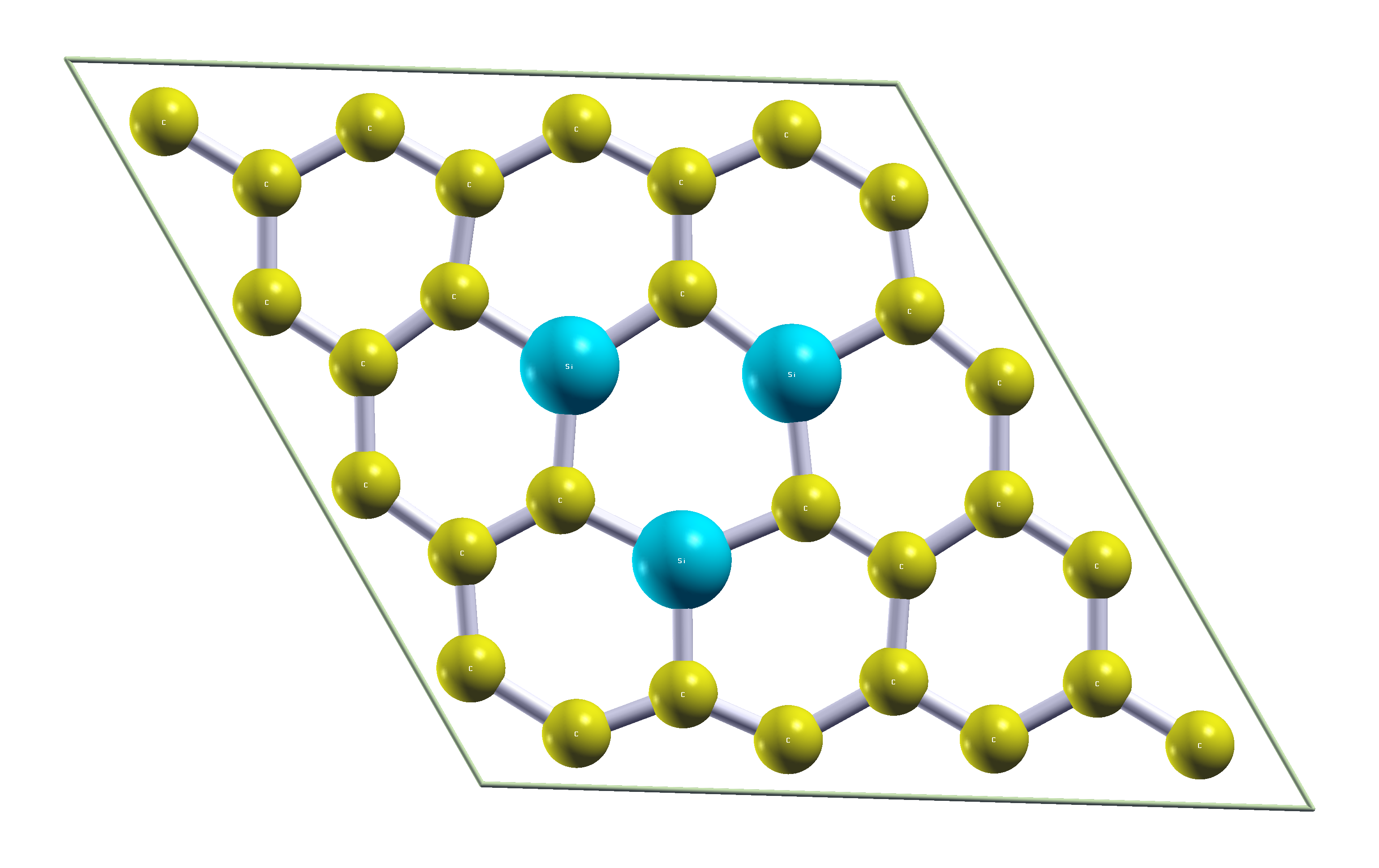}
   \includegraphics[width=0.23\textwidth]{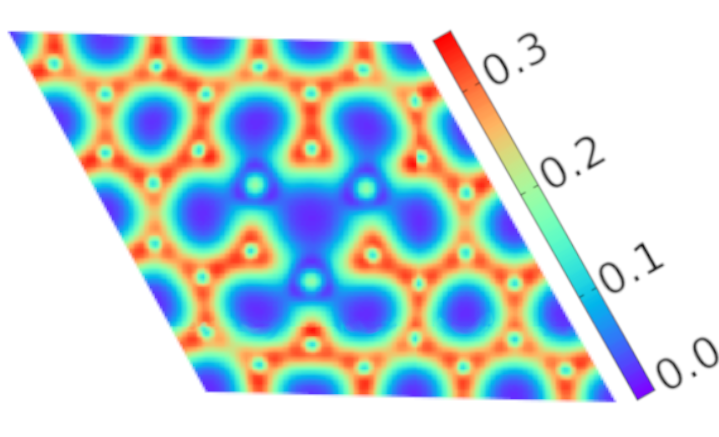}\\
     \includegraphics[width=0.23\textwidth]{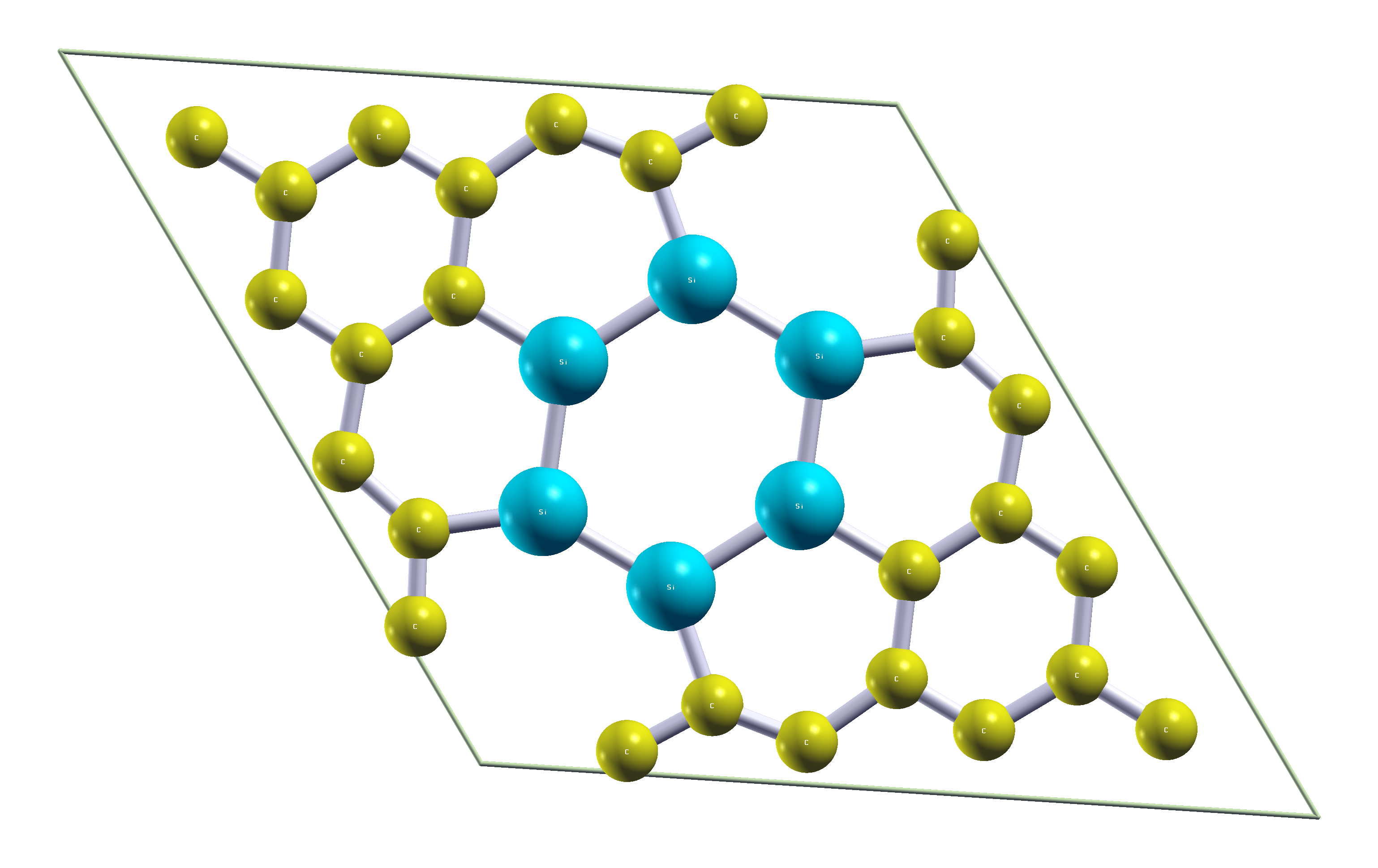}
  \includegraphics[width=0.23\textwidth]{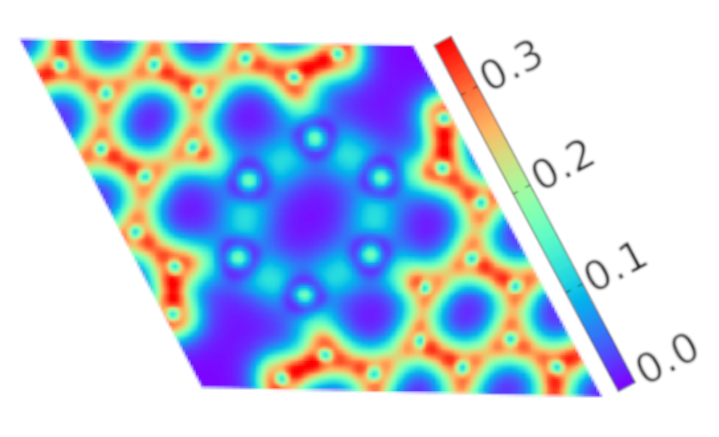}
 \caption{A graphene nanosheet with a triangle configuration of Si-doping atoms (top left panel)
         and its charge density distribution (top right panel). 
         Two Si-atoms are located at the ortho-positions
         while one Si-atom is at the para-position.
         The graphene nanosheet with the ``dot'' configuration of Si-doping atoms (bottom left panel) 
         and its charge density distribution (bottom right panel). 
         The six Si-atoms are distributed over the ortho-, meta-, para-positions.}
\label{fig04}
\end{figure}

We now consider Si atoms substitutionally doped in the graphene nanosheet with different 
geometries or configurations: triangle- and ``dot''-shapes. In the triangle Si-doped graphene, 
we assume three Si atoms (blue color) forming a triangle shape embedded in the center of the graphene nanosheet 
as is shown in \fig{fig04} (top left panel). Two of the Si-atoms are placed at the ortho-positions 
and the third one is embedded in a para-position forming a triangle shape.
In addition, a configuration with six Si-atoms forming a ``circle'' or a ``quantum dot'' shape embedded in the 
center of the graphene nanosheet (bottom left panel) is also considered in this study in which 
the six Si-atoms are distributed over the ortho-, meta-, para-positions.

The configuration and the distribution of embedded Si atoms in graphene have been investigated \cite{Wei2014},
and it has been shown that the location of Si atoms on the graphene (not the concentration) can easily 
be used to tune the electronic structure of the system.
The Si doped graphene nanosheets are called siligraphene nanosheets \cite{Naqvi2018}.
These two selected shapes are analogous to the triangle shape of semiconducting nanowires that have been 
used to control the efficiency of the solar cells \cite{7997887} and the quantum dots embedded in 
semiconductor quantum wires used to design the resulting charge distribution 
\cite{Nzar.25.465302,Nzar_ChinesePhysicsB2016,Nzar_2016_JPCM} and thermoelectric \cite{en12061082} currents. 
Motivated by these geometrical shapes of the semiconducting materials, 
we consider the triangle and dot Si-dopant configurations and 
investigate their electrical and thermal properties.

To check the stability of Si doped graphene we need 
to calculate the formation energy ($E_f$) from the below equation
\begin{equation}
      E_f = E_{\rm T} -  N_{\rm C} \, \mu_{\rm C} -  N_{\rm Si} \, \mu_{\rm Si}.
\end{equation}
Herein, $E_{\rm T}$ is the total energy of the Si doped graphene,  $N_{\rm C}$ and $N_{\rm Si}$ are the number of carbon and silicon atoms in the Si doped graphene, respectively, and $\mu_{\rm C}$ and $\mu_{\rm Si}$ are the chemical potentials of the single carbon and single silicon atom, respectively \cite{PhysRevLett.90.046103,C4NR03247K}. The formation energy of the triangle configuration of Si doped graphene is $-108.701$ eV which is smaller than that of the ``dot'' configuration of the Si doped graphene, $-85.769$ eV. 
The smaller formation energy, the more stable structure is obtained.
So, the triangle configuration of the Si doping atoms is more stable than the dot configuration of the Si doping atoms.

We examined the stability of the triangle configuration of the Si dopant structure
by moving one Si atom from a para-position to a next neighbor site of a para-position and see that the 
formation energy becomes $-106.495$ eV. It indicates that the stability is slightly reduced by moving 
away one Si atom from the triangle configuration at the center of the system.
In addition, if one Si atom of an ortho-position is moved from the dot configuration to a next neighbor 
site of an ortho-position, the formation energy is increased to $-75.095$ eV. 
It demonstrates that our model of a dot configuration is more stable.
It has been reported that if the positions of doped atoms are varied in a structure with low 
doping concentration the stability is slightly changed. But for a high doping concentration, 
changing postions of doping atoms has bigger influences on the stability of the structure \cite{Rani2014}.

The charge density distribution of the graphene nanosheet with the triangle (top right panel) and 
the ``dot'' shape (bottom right panel) 
are demonstrated in \fig{fig04}. We should mention that the Si-atoms embedded in the graphene changes 
the bond length C-C to $1.413$ ${\angstrom}$ in the triangle, and to $1.53$ ${\angstrom}$ in the ``dot'' structures. 
The bond length modification of C-C can be referred back to the repulsion force generated 
between C and Si atoms. These changes influence the charge distribution of the system.
It can be clearly seen that in both structures the charge is delocalized around the Si-atoms. This indicates that 
the Si-atoms loose charge and act as ``donors'' \cite{MIAO2017111}.
The delocalization of charge around the Si-atoms may also be referred to the 
fact that the Si-atoms have a larger atomic radius than carbon \cite{Houmad2015}. 
The delocalization of charge has also been observed in other materials such as semiconductors 
\cite{PhysicaE.64.254} leading to enhanced transport. 
The strength of the interaction is thus increased represented by a repulsion force that 
expels charge away from the center of the siligraphene nanosheet, i.\ e., 
charge carriers exceed a bit in other places of the system.

\begin{figure}[htb]
\centering
    \includegraphics[width=0.35\textwidth]{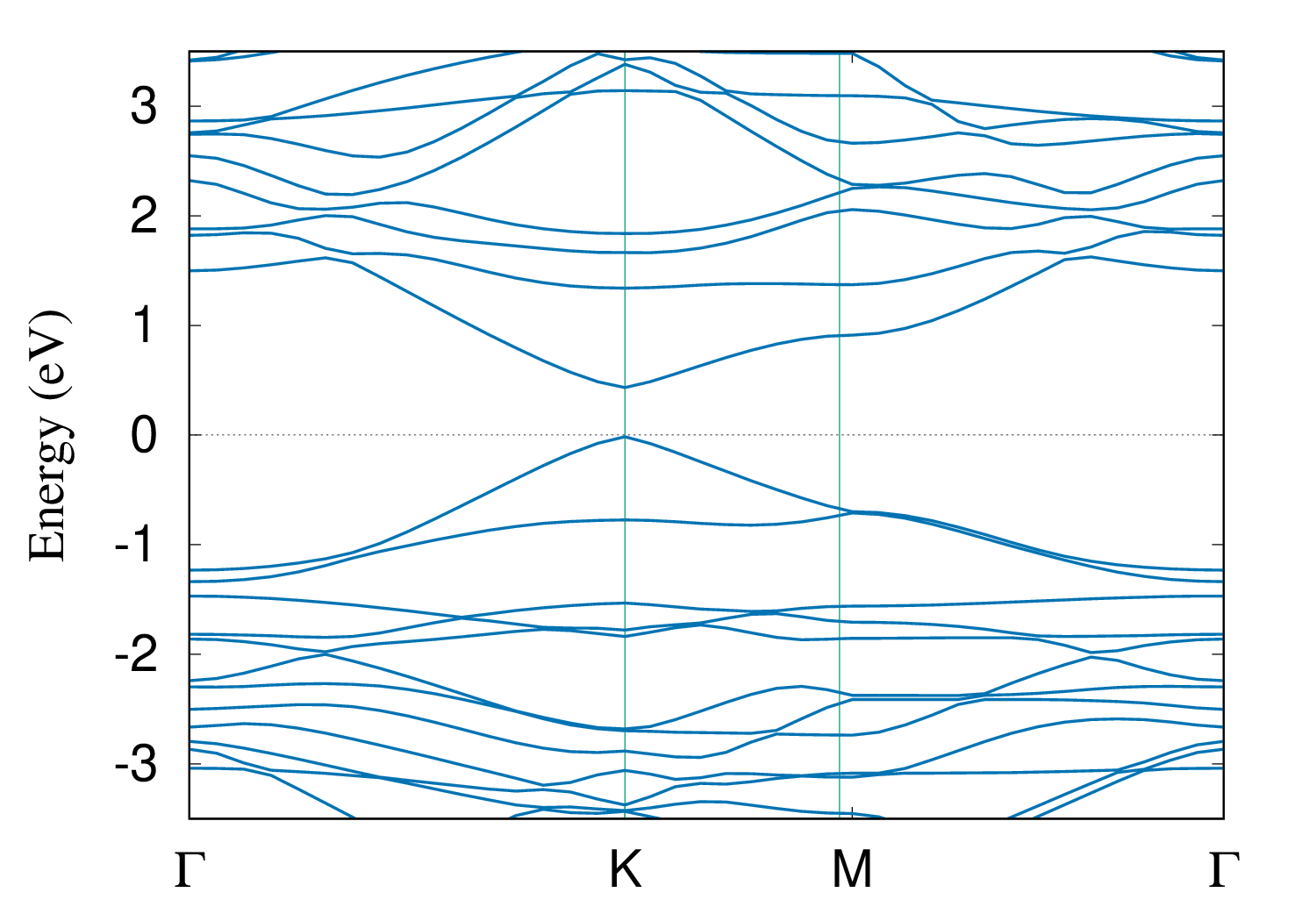}\\
    \includegraphics[width=0.35\textwidth]{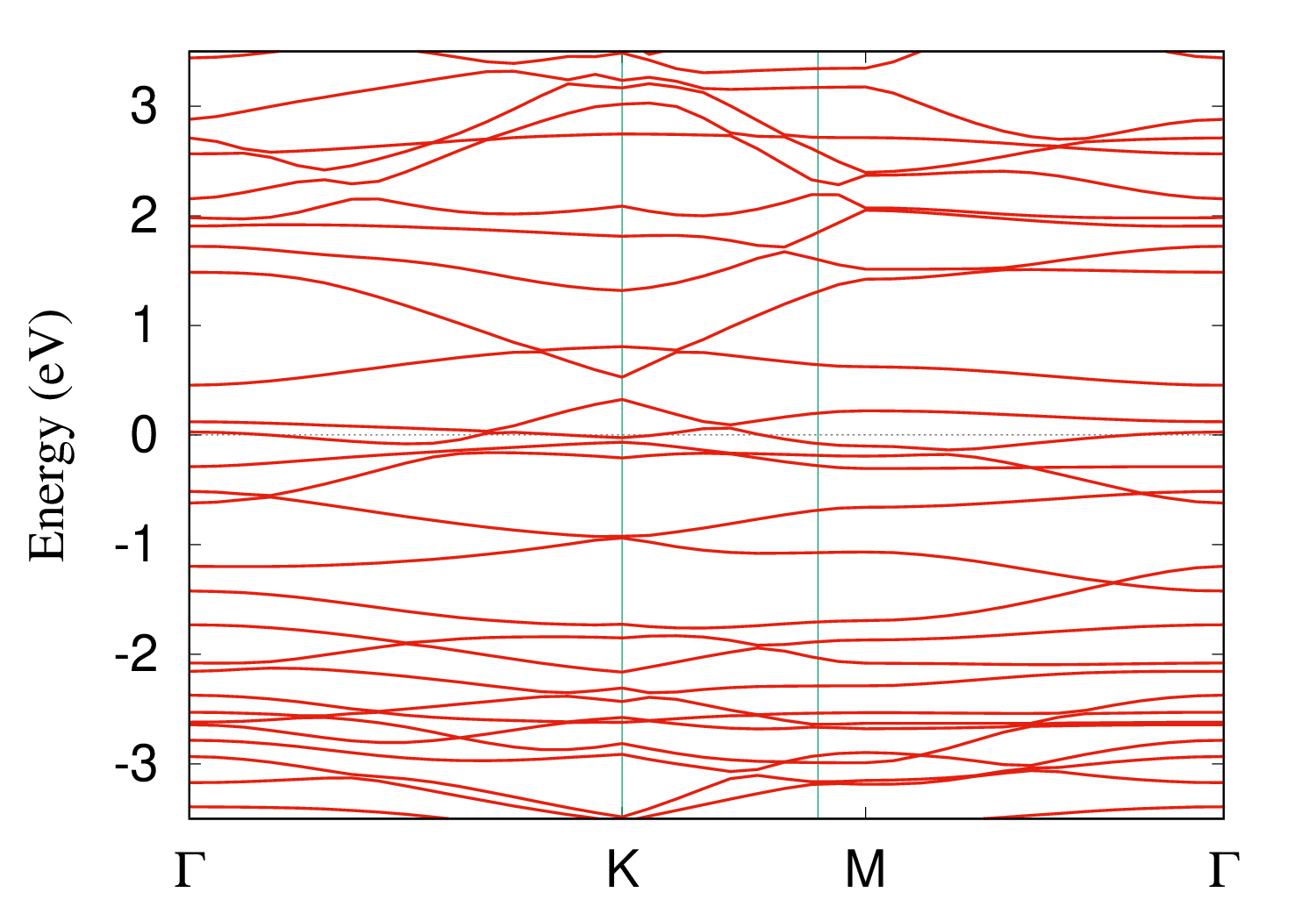}
 \caption{Calculated electronic band structure of the graphene nanosheet with triangle (top panel) 
         and ``dot'' shape (bottom panel) Si-dopants. The horizontal dashed black lines represent 
         the Fermi level, $E_{F}$.}
\label{fig05}
\end{figure}

Figure \ref{fig05} shows the electronic band structure of the triangle (top panel) and the
``dot'' (bottom panel) Si-doped graphene nanosheet.
The repulsion force formed in the presence of triangle Si-dopant opens a band gap, $E_g = 0.448$ eV, at 
the $K$ point as is seen in \fig{fig05} (top panel) \cite{Sahu2017,Shi2015}. This leads to the Si-doped graphene becoming a semiconducting material.

The electronic band structure of the ``dot'' Si-doped graphene gives a totally different physical picture 
in which valence bands cross the Fermi energy. The crossing energy bands have been predicted for different 
material structures and it has been demonstrated that the magnitude and the direction of the energy band 
gap can be sensitively controlled by the dopant type and concentration \cite{Tsai2016,Zollner_2018}. 
The crossing band structure is related to the atomic concentration 
in the primitive unit cell of the Si-doped graphene nanosheet in which the atoms feel almost 
the same potential energy at a specific ratio of dopant concentration. Consequently, the band gap 
closes down, and the conduction, or the particulars of the valence band crossing the Fermi level, depend 
on the dopant type \cite{MIAO2017111, Zollner_2018}.

\begin{figure}[htb]
\centering
    \includegraphics[width=0.38\textwidth]{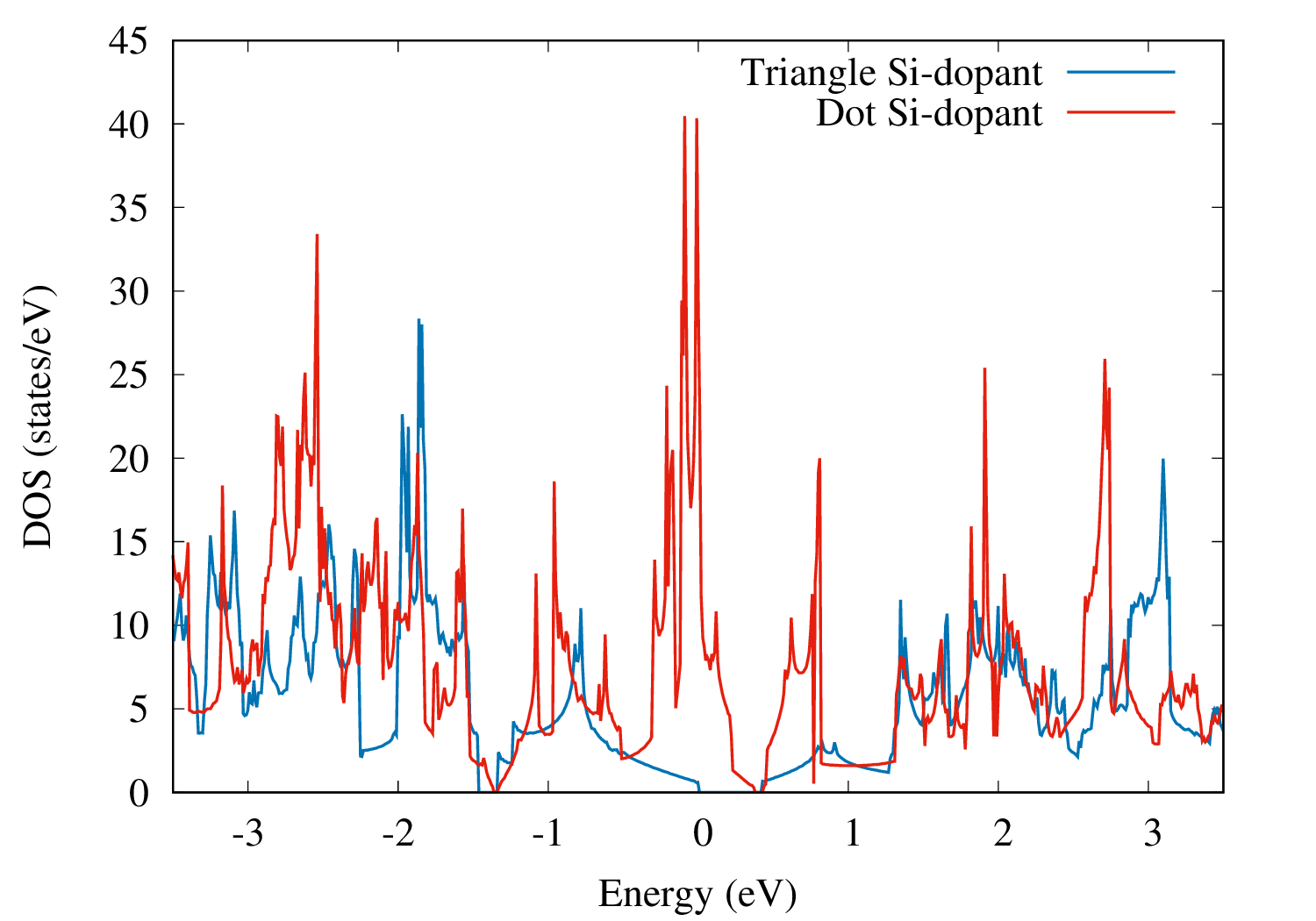}
 \caption{Density of state, DOS, of the triangle (blue color) and ``dot'' (red color) 
         Si-doped graphene nanosheets.}
\label{fig06}
\end{figure}

The changes in the band structure will directly influence the DOS as is displayed in \fig{fig06} for the 
graphene with triangle (blue color) and ``dot'' (red color) Si-dopant.
For instance, the DOS vanishes in the range of the band gap for the triangle Si-doped graphene, and
the DOS around the Fermi level is increased to finite values for the ``dot'' Si-doped graphene
due to the formation of the Fermi-momentum states. 

\begin{figure}[htb]
\centering
   \includegraphics[width=0.35\textwidth]{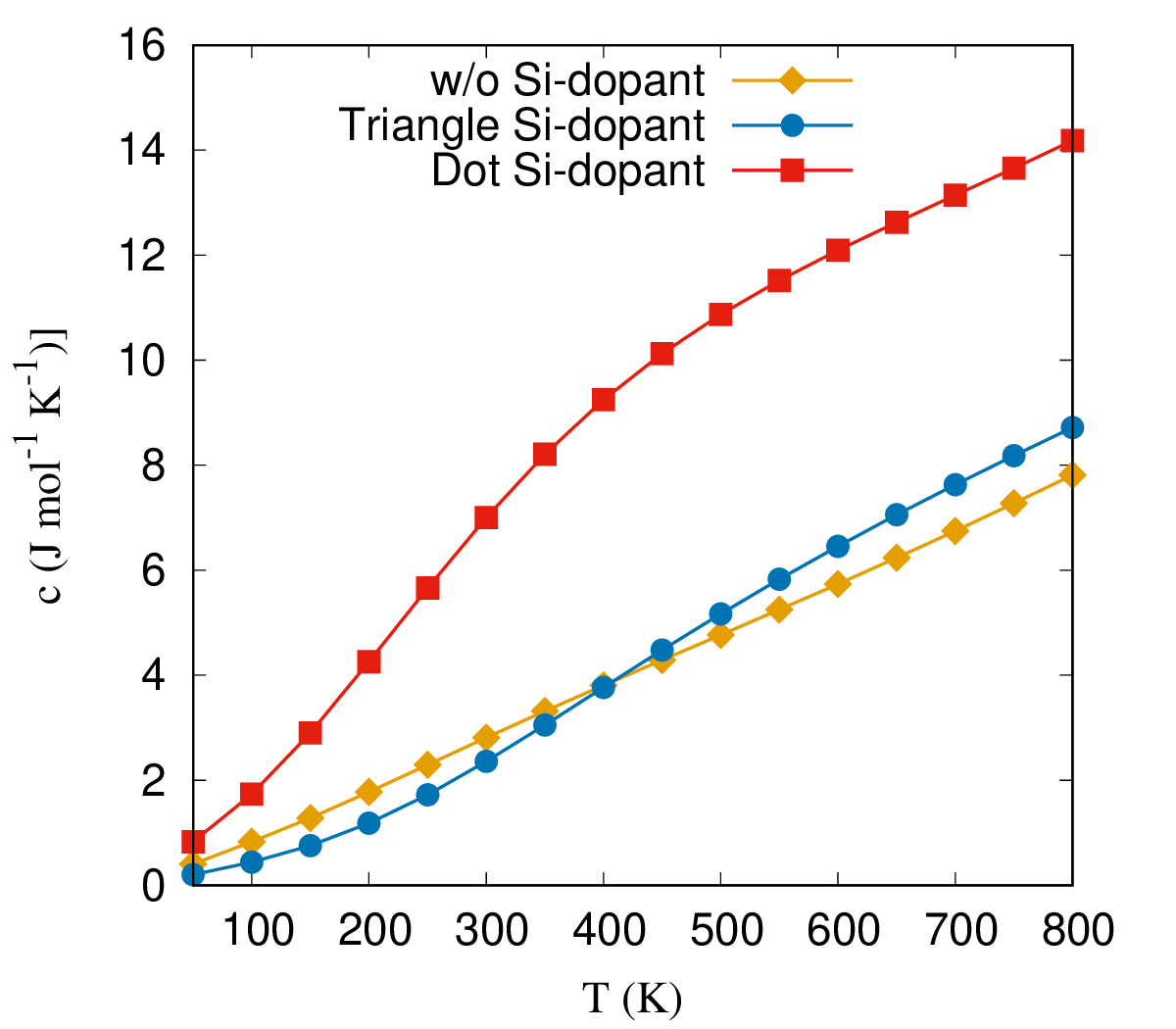}
 \caption{Specific heat, $c$, for the graphene without (w/o) Si-dopant (golden diamond), and 
the graphene with triangle (blue circle) and dot (red square) Si-dopant.}
\label{fig07}
\end{figure}

We now present the thermal properties of the aforementioned structures including the 
specific heat and the electronic thermal conductivity. The Heat capacity can be calculated 
or measured by the ratio of the heat added to or removed from the graphene or Si-doped graphene to 
the resulting temperature change \cite{Wang2014}. 
Therefore, the specific heat can be defined as the heat capacity per unit mass of the material. 
Figure \ref{fig07} shows the specific heat, $c$, for the graphene without (w/o) Si-dopant (golden diamond), 
and the graphene with triangle (blue circle) and ``dot'' (red square) Si-dopant. 
It can be seen that the specific heat increases with the temperature for all 
three cases. The specific heat for the triangle Si-doped graphene is competing with the pristine graphene structure.
Below $400$~K the specific heat of the triangle Si-dopant structure is decreased due to 
the opening band gap that resists the heat transport at ``low'' temperature. 
But above $400$~K the the specific heat is increased compared to the pristine graphene which is related to the 
effect of high temperature reducing the effective band gap.
Furthermore, the specific heat is drastically enhanced for the ``dot'' Si-doped graphene in which 
valence bands are crossed by the Fermi energy. The crossing valence bands arise a heat transfer in the system.
We should mention that our calculated specific heat for the 
pristine graphene is in a good agreement with other calculations of the specific heat for graphene 
valid for temperatures below $800$ K \cite{Wang2017}.

\begin{figure}[htb]
\centering
    \includegraphics[width=0.35\textwidth]{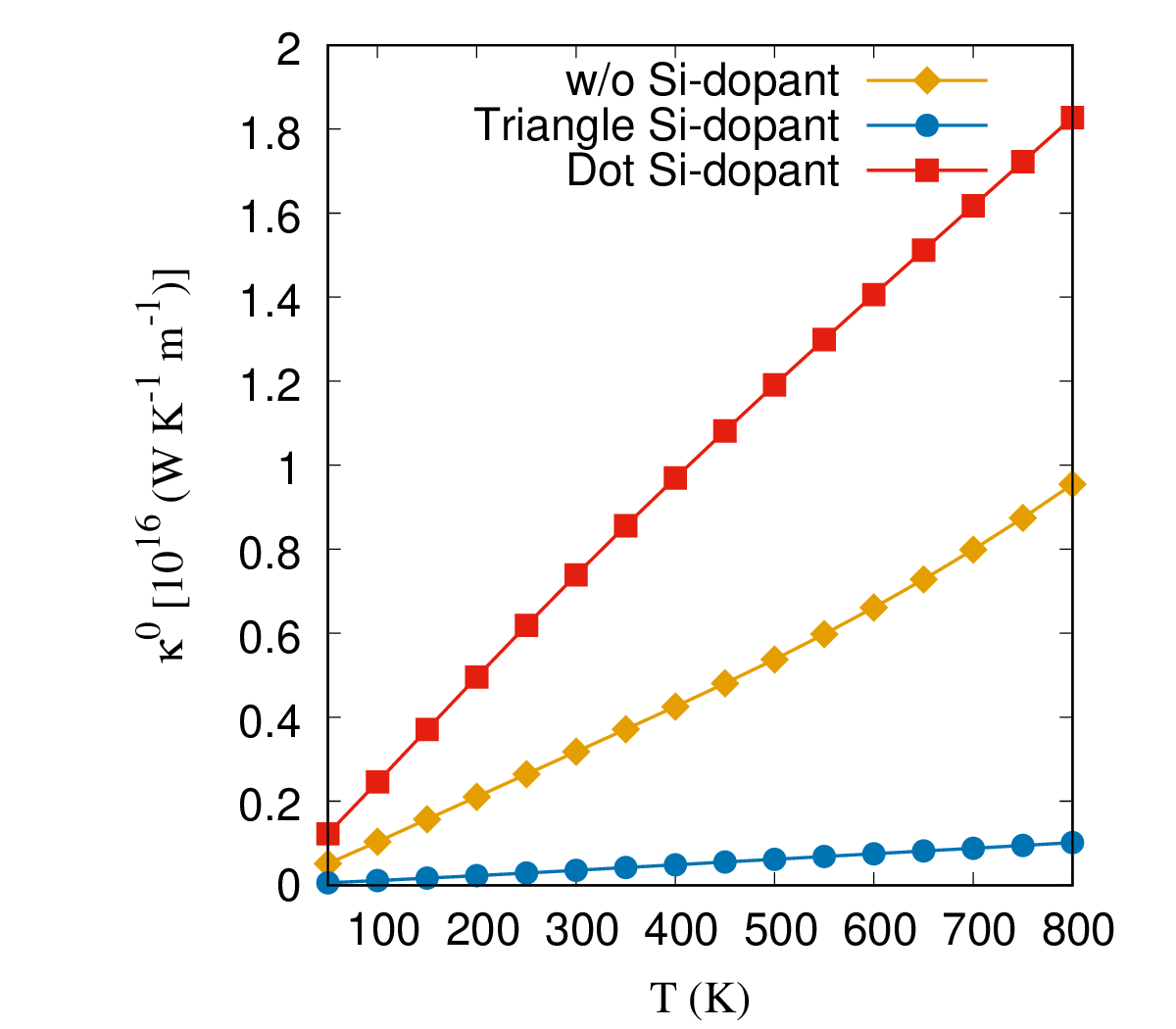}
 \caption{Thermal conductivity, $\kappa^0$, for the pristine graphene (golden diamond), and 
         the graphene with triangle (blue circle) and ``dot'' (red square) Si-dopant.}
\label{fig08}
\end{figure}

In \fig{fig08} the electronic thermal conductivity is presented for the pristine graphene (golden diamond), and 
the graphene with triangle (blue circle) and ``dot'' (red square) Si-doping atoms.
The opening of the band gap for the graphene with triangle Si-doping atoms arises less DOS around the Fermi energy.
Consequently, the charge carriers is decreased in the selected range of temperatures and the thermal 
conductivity is thus suppressed compared to the pristine graphene.
But the increased DOS in the graphene with dot Si-dopant configuration increases the number of 
charge carriers and the thermal conductivity is thus enhanced.\\

\section{Conclusions}
\label{Sec:Conclusion}

We have studied thermal properties of graphene nanosheets with Si atoms with triangle or ``dot'' shape 
configuration of dopants. Density functional theory had been used to calculate the 
band structure, the density of states, and the charge density distribution.
We found that the Si-impurities in both doped systems play a donor role giving charge to the 
graphene structure leading to delocalized charge around the Si-impurities.
As a result, a repulsion force in the triangle Si-doped graphene arises a band gap that 
can be controlled by the concentration of the Si-dopant.
The opening band gap in the triangle Si-doped graphene 
leads to decreases in both the specific heat and the thermal conductivity at low temperatures.
Furthermore, as the concentration of silicon atoms is higher in the ``dot'' Si-doped 
graphene the induce repulsion forces are higher than in the triangular configuration of dopants.
Valence bands thus cross the Fermi-energy in 
the ``dot'' Si-doped graphene with resulting increases in the density of state and the 
number of charge carriers. As a result, the specific heat and the thermal conductivity are enhanced.

\begin{acknowledgments}
This work was financially supported by the University of Sulaimani and 
the Research center of Komar University of Science and Technology.
The computations were performed on resources provided by the Division of Computational Nanoscience 
at the University of Sulaimani.
\end{acknowledgments}


%

\end{document}